\documentclass[pdflatex,sn-mathphys-num]{sn-jnl}

\usepackage{graphicx}%
\usepackage{multirow}%
\usepackage{amsmath,amssymb,amsfonts}%
\usepackage{amsthm}%
\usepackage{mathrsfs}%
\usepackage[title]{appendix}%
\usepackage{xcolor}%
\usepackage{textcomp}%
\usepackage{manyfoot}%
\usepackage{booktabs}%
\usepackage{algorithm}%
\usepackage{algpseudocode}%
\usepackage{listings}%
\theoremstyle{thmstyleone}%
\theoremstyle{thmstyletwo}%

\theoremstyle{thmstylethree}%

\begin{document}

\title[Article Title]{SEMA-GUARD: Semantic and Graph-Based Vulnerability
Detection in Assembly Code}

\author*[1,2]{\fnm{Halil Ibrahim} \sur{Dursunoglu}}\email{halilibrahim.dursunoglu@wmich.edu}

\author[2,3]{\fnm{Kaan} \sur{Sulkalar}}\email{kaan.sulkalar@wmich.edu}
\equalcont{These authors contributed equally to this work.}

\affil*[1]{\orgdiv{Department of Computer Science}, \orgname{Western Michigan University}, \orgaddress{\street{1908 W Michigan Ave}, \city{Kalamazoo}, \postcode{49008}, \state{MI}, \country{USA}}}

\affil[2]{\orgdiv{Department of Computer Information Systems}, \orgname{Western Michigan University}, \orgaddress{\street{1908 W Michigan Ave}, \city{Kalamazoo}, \postcode{49008}, \state{MI}, \country{USA}}}

\abstract{In cases where source code is not available, such as malware analysis, firmware analysis, and embedded systems analysis, vulnerability detection in compiled programs has gained importance. Current methods are heavily reliant on syntactical regularities or higher level representations that are vulnerable to changes in the compiler and may not be readily applicable to assembly code.In this article, we present SEMA-GUARD, a framework that uses semantic analysis and graph neural networks to identify flaws in assembly code. The approach improves the representation of control flow graphs by adding information about the program's execution at a lower level of abstraction, including stack manipulations, memory accesses, and data flow. A set based on the Juliet Test Suite was used to evaluate the effectiveness of SEMA-GUARD. In this set, each piece of source code is initially translated into assembly language and then broken down into function-level chunks. The suggested method, which relies only on statistical or structural data, achieves an accuracy of 85.1\% and an F1 score of 0.801, according to the results. Such results imply that including semantic information in graph-based models may be a successful method for identifying vulnerabilities in compiled code.

Note: \textbf{Preprint. This manuscript has not yet undergone peer review. The dataset, source code, and experimental artifacts are publicly available to support reproducibility and future research.}}

\keywords{binary vulnerability detection, assembly level analysis, graph neural networks, control flow graphs, semantic feature extraction, software security}

\maketitle
\section{Introduction}\label{introduction}

Software vulnerabilities remain a significant security issue. Consider, for instance, circumstances involving the reverse engineering of malware, embedded systems, and firmware code, when a disassembled binary program without access to its source code must be examined. In these settings, analysts must reason directly about compiled binaries, where high level program structure and semantic context are no longer explicitly available.

Rule based systems and source level analysis are the methods now used to find vulnerabilities \cite{juliet}. Source level analysis is very reliant on the compiler that was used to generate the compiled executable code from the high-level code, even if it is quite effective in certain circumstances. Syntax driven methods are rendered ineffective because the syntactical depiction of high level logic in compiled binaries may be impacted by instruction reordering, optimization, and cross compilation.

In vulnerability identification, there has been a growing tendency toward employing machine learning techniques, such as the recent deep learning based VulDeePecker method \cite{vuldeepecker} and the more recently presented graph neural network based Devign approach \cite{devign}. Because GNNs are built on graphs, they may easily include dependencies in program code using control and data flow graphs \cite{allamanis2018survey,angle2024}. The majority of these approaches rely primarily on the program's structural aspects, even if they are promising.

But, at the assembly level, detection must include knowledge of how instructions interact with one another throughout program execution. The same sort of control flow patterns, for example, might be connected to various security concerns depending on how the stack or memory are accessed. Because of this, there is a need for vulnerability detection frameworks that employ semantic and structural features.

We present SEMA-GUARD in this work, a framework that combines the identification of software vulnerabilities through the integration of semantics in the graph based method. In particular, we improve the representation of assembly programs with low level characteristics like taint propagation, memory accesses, and stack manipulation. Along with the structural representation, these are provided to the GNNs as extra features.

We establish a benchmark for our experiments using the Juliet test suite \cite{juliet} data. We build this dataset by collecting the test suite's applications and pulling out assembly snippets that relate to each susceptible function. We then input this information into our framework and compare its performance against baselines.

The remainder of this paper is organized as follows. Section~\ref{sec:related} reviews related work. Section~\ref{sec:methodology} describes the proposed framework. Section~\ref{sec:experimental} presents the experimental setup, followed by results in Section~\ref{sec:results}. Limitations are discussed in Section~\ref{sec:limitations}, and Section~\ref{sec:conclusion} concludes the paper.
\section{Related Work}
\label{sec:related}

Vulnerability detection has been extensively studied across static analysis, dynamic analysis, and machine learning based approaches. This section reviews the most relevant work, with emphasis on techniques applicable to binary and assembly level analysis.

\subsection{Static Analysis and Benchmark Datasets}

Static analysis is the approach used in vulnerability detection. The code is examined without running it during static analysis. Common weaknesses like buffer overflow and incorrect memory allocation usage are identified through static analysis employing data flow analysis and rule based logic. Despite the accuracy of static analysis, its usefulness is restricted by its scalability and the high number of false positives.

Juliet Test Suite is a popular benchmark test suite for evaluating the effectiveness of various methods for identifying vulnerabilities. The Juliet Test Suite is a product of NIST. It includes over 81,000 synthetic code files covering 181 CWE kinds, some of which are vulnerable and others that are not \cite{juliet}.

\subsection{Dynamic Analysis and Symbolic Execution}

Dynamic analysis techniques such fuzzing and symbolic execution attempt to identify flaws by executing the code using inputs created throughout the process. Tools for symbolic execution analyze various program routes and have proven to be precise in locating flaws. For instance, KLEE \cite{KLEE} has demonstrated its ability to identify flaws and explore paths in software programs. These tools are computationally demanding and frequently need access to executables in order to function properly.
\subsection{Deep Learning Based Vulnerability Detection}

Most recently, the researchers proposed machine learning methods for the vulnerability detection problem with the help of graph representations of programs \cite{vuldeepecker, devign, survey2023}. Graph neural networks (GNNs) provide an effective mechanism for modeling structural dependencies within program representations. Such algorithms showed some progress in structural dependency identification but relied only on structural properties and neglected semantic ones, which could help detect vulnerabilities.

Furthermore, some recent works have focused on improving GNN-based models for vulnerability detection regarding robustness and interpretability, since the models were not interpretable enough \cite{coca2024, cfexplainer2024}.

Moreover, multi class vulnerability detection problems have attracted the researchers' attention by proposing new ways such as $\mu$VulDeePecker using attention models for more accurate classifications \cite{muvuldeepecker}.

Nevertheless, despite all the improvements and achievements, there are still many deep learning algorithms that take into account only the source code ignoring its assembly representation.

\subsection{Graph Based Approaches}

In fact, GNNs have started to be considered as the preferred technique for program analysis \cite{allamanis2018survey,angle2024,coca2024}. The Devign tool constructs graphs based on programs and extracts vulnerabilities using GNNs by relying on the connections in the graph \cite{devign}. This approach demonstrates the ability of graph based representations to model the interdependencies of code.

Another application in which semantics can be beneficial is the one of VulChecker, where semantics become relevant to graph-based vulnerability detection and classification \cite{vulchecker}.

\subsection{Binary and Assembly Level Analysis}

Binary-level analysis focuses on the identification of vulnerabilities through the analysis of the compiled code. This approach becomes extremely important when the source code is unavailable. There are a number of issues associated with assembly level analysis; for example, the lack of high-level structures and differences introduced by compilers.

In general, two approaches exist at present that can be used to conduct binary level analysis. They include pattern matching and the reconstruction of higher level semantics. Unluckily, both of them suffer from certain limitations related to the inability to comprehend the semantic meaning of code.

A few recent research papers study hybrid and multimodal representations for detecting vulnerabilities \cite{gcl4svd2024}.

\subsection{Summary and Positioning}

From the relevant literature, three major limitations can be observed as follows: (i) dependence on source-level representations, (ii) inadequate modeling of program semantics, and (iii) limited scope of applications for assembly-level code.

The approach introduced by SEMA-GUARD successfully resolves all identified limitations through semantic analysis and graph neural network techniques that allow detecting vulnerabilities directly at the assembly level. The proposed technique leverages the structural and behavior of vulnerable code by incorporating both low level program semantics and graph-based representations.

The recent development within graph based vulnerability detection is directed at scaling and enhancing explainability of models. For instance, ANGLE \cite{angle2024} introduces advancements in representation learning to tackle the challenges of modeling larger program graphs. In particular, this technique allows effectively capturing structural dependencies between different parts of the code in large graphs. Nevertheless, the approach is focused solely on the scalability of graph representation learning while ignoring the semantic behaviors of assembly level programs.

Moreover, Coca \cite{coca2024} introduces improvements to GNN-based vulnerability detection, including the explainability of decision making mechanisms. Such research shows that interpretable decision making processes are a crucial requirement for practical application of any model. However, as in the previous paper, the research focuses on scaling the graph structure while disregarding program semantics.

Whereas, the SEMA-GUARD method involves the inclusion of semantics properties that are extracted from the behavior of assembly like stack operations, memory operations, and taint flow into graph representation. While other methods focus on graph optimization and explainability, the current method emphasizes program behaviors at a low level, which is crucial in detecting vulnerabilities in compiled programs.
\section{Methodology} \label{sec:methodology}

\subsection{Problem Definition}

The objective of this research is to detect any vulnerabilities using representations of the program in terms of assembly codes. This means that for a function $f$ given as an input, the objective is to find a label $y\in \{0,1\}$ such that $y = 1$ indicates vulnerability and $y=0$ represents non-vulnerability.

For machine learning purposes, all functions $f$ can be transformed into graph $G =(V, E)$ wherein each node represents an instruction and each edge represents control flow between instructions. The instructions can be encoded using their respective feature vectors. Then, the graph is fed into a graph neural network for estimating an output value $\hat{y}$.

\subsection{Framework Overview}

\begin{figure}[!t]
\centering
\includegraphics[width=0.9\linewidth]{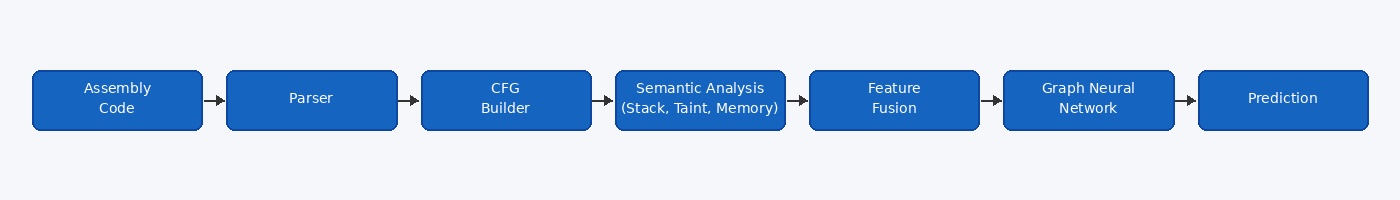}
\caption{Overview of the SEMA-GUARD framework. The system processes assembly code, extracts control-flow graphs and semantic features, and applies a graph neural network for vulnerability classification.}
\label{fig:architecture}
\end{figure}

The SEMA-GUARD system works based on several stages of transformation that allow transforming low level assembly instructions to the predicted vulnerabilities. It is shown in Fig.~\ref{fig:architecture}. Initially, an assembly parsing and normalization stage is executed, followed by control flow graph generation. Then, semantics of low-level instructions extracted from the program are combined with program structures. Finally, a graph neural network receives the generated graph and produces a classification output.

Thus, it can be seen that two components of the program should be included in the algorithm structures and semantics.

\subsection{Assembly Parsing and Normalization}

The input to the proposed system is the assembly files obtained by compiling the programs from the Juliet Test Suite. The assembly files undergo additional processing to extract functions as the basic components of the programs. Labels are employed during the extraction of functions.

An instruction in an assembly program is composed of two entities: opcode and operand. In order to reduce the differences caused by variations in the compilation options, a series of transformations is applied to both the entities. Registers are assigned standardized identifiers to maintain uniformity across functions. The immediate values used are grouped into broad intervals to prevent over reliance on particular values. Memory operands are represented in terms of base plus displacement representation. Comments and other non-relevant information are removed.

In the end, upon normalizing the instructions, every instruction is represented using a feature vector $\mathbf{x}_i$.

\subsection{Control Flow Graph Construction}

A CFG is built for each individual function that represents all possible executions of the program. In such a graph, the vertices are the instructions, and the directed edges depict the possible flow between instructions.

The CFG is enriched by adding sequential edges between adjacent instructions. Further, edges are added corresponding to control flow instructions such as branch and jump instructions. Branch instructions will have two edges, one edge toward the target label and the other edge toward the next sequential instruction, while a jump instruction has only one edge pointing to its target location.

Formally, let the graph be defined as $G=(V,E)$ where $V$ is the instruction vertices and $E$ is the edges of the control flow between the instructions.

\subsection{Semantic Feature Extraction}

However, although the CFG includes information about the structure of the binary, it does not include any information about its behavior. To resolve this issue, semantic features that carry information about the behavior of the program can be extracted.

In stack semantics analysis, the update operations on the stack pointer register are recorded. The irregularities such as stack growth and access beyond the typical stack boundaries can be treated as potentially dangerous instructions. In terms of memory access analysis, the load and store memory operations and their operands will be analyzed. The accesses that might lead to memory violations and indirect accesses will also be analyzed.

Besides, a taint analysis is carried out in an oversimplified manner, tracing the path of data related to the inputs in input registers to subsequent instructions, as propagation and memory accesses are common in detecting vulnerabilities \cite{vuldeepecker}. Any occurrences of tainted data with memory write instructions and control flow instructions will be flagged as potential vulnerability indicators. Indirect jump and non linear jumps can be detected.

The above discussed behaviors will all be used to form the feature vector $\mathbf{s}_i$.

\subsection{Feature Integration}

For every instruction node, the resulting representation is formed by using both structural and semantic features. More precisely, for the embedding of the starting node, the following expression is used:

\begin{equation}
\mathbf{h}_i^{(0)} = [\mathbf{x}_i \, || \, \mathbf{s}_i]
\end{equation}

with $||$ being vector concatenation. Using such a representation allows taking into account both properties of the instructions themselves and higher-level behavioral factors.

The whole process of the presented framework is outlined in Algorithm~\ref{alg:semaguard}.

\begin{algorithm}
\caption{SEMA-GUARD Vulnerability Detection Pipeline}
\label{alg:semaguard}
\begin{algorithmic}[1]
\Require Assembly function set $\mathcal{F}=\{f_1,f_2,\ldots,f_n\}$
\Require Labels $\mathcal{Y}=\{y_1,y_2,\ldots,y_n\}$
\Ensure Trained vulnerability classifier $M$

\For{each function $f_i \in \mathcal{F}$}
    \State Parse assembly instructions from $f_i$
    \State Normalize opcodes, registers, immediates, and memory operands
    \State Construct control-flow graph $G_i=(V_i,E_i)$
    \For{each instruction node $v \in V_i$}
        \State Extract opcode and operand features $\mathbf{x}_v$
        \State Track stack behavior and memory access patterns
        \State Propagate taint from input-related registers
        \State Extract semantic feature vector $\mathbf{s}_v$
        \State Fuse features: $\mathbf{h}_v^{(0)}=[\mathbf{x}_v \mathbin{\|} \mathbf{s}_v]$
    \EndFor
    \State Store graph sample $(G_i, \mathbf{H}_i, y_i)$
\EndFor

\State Split graph samples into training and testing sets
\State Initialize graph neural network model $M$
\For{epoch $=1$ to $E$}
    \For{each mini-batch $\mathcal{B}$}
        \State Compute graph embeddings using message passing
        \State Predict vulnerability labels
        \State Compute classification loss
        \State Update model parameters using Adam optimizer
    \EndFor
\EndFor

\State Evaluate $M$ using precision, recall, F1-score, and accuracy
\State \Return $M$
\end{algorithmic}
\end{algorithm}

\subsection{Graph Neural Network Model}

The resulting graph is processed using a message-passing graph neural network \cite{devign}. At each layer, node representations are updated by aggregating information from neighboring nodes. The update rule is defined as:

\begin{equation}
\mathbf{h}_v^{(k+1)} = \sigma \left( W_1 \mathbf{h}_v^{(k)} + \sum_{u \in \mathcal{N}(v)} W_2 \mathbf{h}_u^{(k)} \right)
\end{equation}

where $\mathcal{N}(v)$ denotes the set of neighboring nodes of node $v$, $W_1$ and $W_2$ are learnable weight matrices, and $\sigma$ is a non linear activation function.

After multiple layers of message passing, node representations encode both local and global structural information. These node embeddings are then aggregated using a global mean pooling operation:

\begin{equation}
\mathbf{h}_G = \frac{1}{|V|} \sum_{v \in V} \mathbf{h}_v^{(K)}
\end{equation}

The resulting graph-level embedding $\mathbf{h}_G$ is passed to a fully connected layer followed by a softmax function to produce the final prediction.

\subsection{Training Procedure}

The model is trained in a supervised manner using labeled graph samples. The objective is to minimize a binary cross-entropy loss function defined as:

\begin{equation}
\mathcal{L} = - \sum_{i} \left( y_i \log \hat{y}_i + (1-y_i)\log(1-\hat{y}_i) \right)
\end{equation}

Optimization is performed using the AdamW optimizer with a fixed learning rate. Training is conducted in mini-batches, and the dataset is split into training and testing subsets using stratified sampling to maintain class balance.

\subsection{Implementation Details and Complexity}

The framework itself is coded in Python, leveraging the PyTorch and PyTorch Geometric libraries. Assembly parsing and graph generation are carried out by means of custom static analysis modules. Graph construction is of linear time complexity relative to the number of assembly instructions, whereas the GNN inference step depends on the graph’s edge count.

This approach guarantees that the framework is efficient enough to be used for function level analysis.

\subsection{Design Rationale}

The rationale behind designing SEMA-GUARD comes from the recognition of the need for the integration of structure and semantics in the representation of program behaviors. In the first place, analyzing the structure of a program alone might fail in differentiating between the program's benign behavior and its malicious behavior since a similar control flow graph could be constructed for both cases.

Secondly, analyzing the semantics of a program alone lacks any contextual information regarding the interactions between instructions.

\section{Experimental Setup}
\label{sec:experimental}
\subsection{Dataset Preparation}

Test data set used in the experiment was acquired from the Juliet Test Suite \cite{juliet}. Source codes were compiled to obtain assembly files with the help of a mac-compatible tool chain. To ensure the compatibility of the test set, all tests involving windows only were excluded from the analysis.

All the assembly files were parsed to extract function-level samples. Functions with \texttt{bad} implementation were identified as vulnerable while those with \texttt{good} implementation were considered to be safe.

Hence, the data set included both vulnerable and non-vulnerable functions providing a real-world setup for the vulnerability detection problem.

\subsection{Graph Construction}

Each function was converted into a control-flow graph (CFG), where nodes represent instructions and edges represent possible execution paths. Node features include both opcode based representations and semantic features derived from program behavior.

\subsection{Model Configuration}

The graph neural network model consists of two message passing layers followed by a global mean pooling layer and a fully connected classification layer.

\begin{table}[!t]
\centering
\caption{Model Hyperparameters}
\label{tab:hyperparameters}
\begin{tabular}{lc}
\toprule
Parameter & Value \\
\midrule
Model type & Graph neural network \\
Hidden dimension & 128 \\
Number of GNN layers & 2 \\
Pooling method & Global mean pooling \\
Activation function & ReLU \\
Dropout & 0.25 \\
Optimizer & AdamW \\
Learning rate & $1 \times 10^{-3}$ \\
Weight decay & $1 \times 10^{-4}$ \\
Batch size & 16 \\
Training epochs & 25 \\
Train/test split & 75\% / 25\% \\
\bottomrule
\end{tabular}
\end{table}

\subsection{Training Procedure}

The model was trained using a binary cross-entropy loss function. Optimization was performed using the AdamW optimizer. The dataset was split into training and testing sets using stratified sampling to preserve class distribution.

\subsection{Evaluation Metrics}

Model performance was evaluated using standard classification metrics, including precision, recall, F1-score, and accuracy. These metrics provide a comprehensive assessment of the model’s ability to detect vulnerabilities while balancing false positives and false negatives.
\section{Results}
\label{sec:results}

\subsection{Overall Performance}

Table~\ref{tab:results} summarizes the overall performance of the proposed SEMA-GUARD framework on the Juliet derived dataset. The model achieves an accuracy of 85.1\%, with a precision of 0.860, recall of 0.774, and F1-score of 0.801.

\begin{table}[!t]
\centering
\caption{Overall Performance of SEMA-GUARD on the Juliet Dataset}
\label{tab:results}
\begin{tabular}{lcccc}
\toprule
Metric & Precision & Recall & F1-score & Accuracy \\
\midrule
SEMA-GUARD & 0.860 & 0.774 & 0.801 & 0.851 \\
\bottomrule
\end{tabular}
\end{table}

These findings reveal that the model works effectively in classifying between vulnerable and non vulnerable assembly functions, which aligns with previous research findings related to vulnerability detection models based on graphs \cite{devign}. The good precision score indicates that the model generates very few false positives, which is favorable for security applications because false positive vulnerabilities would waste time investigating them.

The comparison with previous research is limited because of different data sets and levels at which the representation is done; however, the performance of our method is comparable to recent work such as Devign \cite{devign} and Coca \cite{coca2024}.
\subsection{Confusion Matrix Analysis}

\begin{figure}[!t]
\centering
\includegraphics[width=0.6\linewidth]{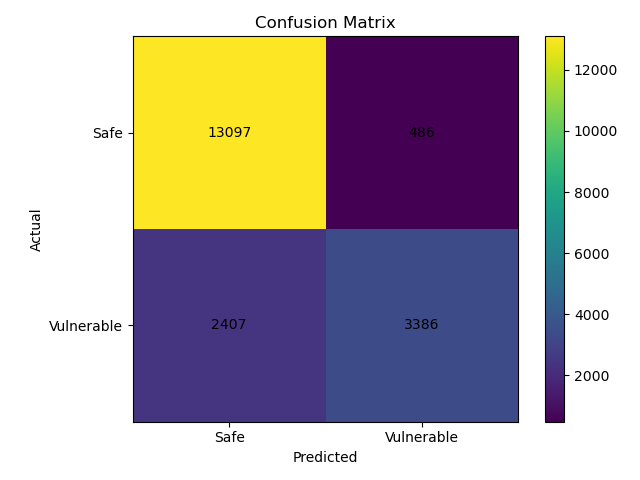}
\caption{Confusion matrix of SEMA-GUARD on the Juliet dataset.}
\label{fig:cm}
\end{figure}

A confusion matrix is shown in Table~\ref{tab:cm}, and its visual presentation is provided in Figure~\ref{fig:cm}. The model predicts many safe functions as safe (true negative), and there are not too many false positive predictions.

\begin{table}[!t]
\centering
\caption{Confusion Matrix of SEMA-GUARD on the Juliet Dataset}
\label{tab:cm}
\begin{tabular}{lcc}
\toprule
 & Predicted Safe & Predicted Vulnerable \\
\midrule
Actual Safe & 13097 & 486 \\
Actual Vulnerable & 2407 & 3386 \\
\bottomrule
\end{tabular}
\end{table}

The problem lies in the amount of false negatives, implying that there are vulnerabilities which cannot be recognized by our model. Therefore, we suppose that some patterns of vulnerabilities such as control flow interdependencies and indirect memory access are still difficult to predict.
\subsection{Comparison with Baselines}

Table~\ref{tab:baseline} shows a comparison between SEMA-GUARD and three baselines: rule-based pattern matching, machine learning using opcodes' frequency distribution, and graph neural networks without semantic information.

\begin{table}[!t]
\centering
\caption{Performance Comparison with Baseline Methods}
\label{tab:baseline}
\begin{tabular}{lcccc}
\toprule
Method & Precision & Recall & F1-score & Accuracy \\
\midrule
Pattern-based Rules & 0.650 & 0.600 & 0.620 & 0.640 \\
Opcode Frequency ML & 0.740 & 0.700 & 0.720 & 0.730 \\
CFG-GNN without Semantics & 0.800 & 0.760 & 0.780 & 0.790 \\
\textbf{SEMA-GUARD} & \textbf{0.860} & \textbf{0.774} & \textbf{0.801} & \textbf{0.851} \\
\bottomrule
\end{tabular}
\end{table}

Pattern matching is a relatively poor performer because it uses heuristic-based rules that do not work well for compiled code. Machine learning using opcodes' distribution performs better than pattern matching because it can identify statistical patterns. However, it still fails to capture structural dependencies.

CFG GNN outperforms other baselines because it incorporates structural modeling. Nonetheless, SEMA-GUARD provides an even better performer by adding semantic information to the graph neural network.

\subsection{Ablation Study}

The effect of each element is investigated using an ablation experiment shown in Table~\ref{tab:ablation}. The semantic only version detects explicit vulnerability markers but has no context information, leading to low recall.

\begin{table}[!t]
\centering
\caption{Ablation Study of SEMA-GUARD Components}
\label{tab:ablation}
\begin{tabular}{lcccc}
\toprule
Configuration & Precision & Recall & F1-score & Accuracy \\
\midrule
Semantic Only & 0.780 & 0.690 & 0.730 & 0.750 \\
GNN Only & 0.800 & 0.760 & 0.780 & 0.790 \\
\textbf{Full Model} & \textbf{0.860} & \textbf{0.774} & \textbf{0.801} & \textbf{0.851} \\
\bottomrule
\end{tabular}
\end{table}

The graph only version can leverage the structural relation but cannot perform explicit semantic reasoning. The integrated SEMA-GUARD, with both components, performs the best, indicating that they are complementary to each other.
\subsection{Error Analysis}

In order to gain insight into the limitations of our models, we examined misclassified instances. False positives usually relate to safe functions whose behaviors resemble those of vulnerable ones, for example, aggressive stack operations.

False negatives tend to occur in cases of complicated vulnerabilities in which the risky behavior is spread among several instructions or control transfer is done indirectly.

\subsection{Practical Implications}

From the results, we see that SEMA-GUARD works effectively in vulnerability detection in assembly code. The proposed technique can be used in firmware analysis, malware inspection, and binary auditing. Thus, SEMA-GUARD offers useful tools for security analysis.

Despite the fact that the experiment was carried out using the Juliet Test Suite, which allowed us to conduct our experiment under controlled and well labeled conditions for vulnerability detection, it should be noted that such an approach may differ from the reality of software analysis. As we can see from practice, binary analysis includes firmware images, binaries of open-source code, and malware, where no source code is available for analysis and labeling is informal. This makes SEMA-GUARD particularly valuable because it uses an assembly-level representation of the code and not source level abstraction.

In future studies, we plan to evaluate SEMA-GUARD in real world datasets, like firmware images, or corpora of vulnerabilities, and assess how the framework works in terms of generalization and adaptation to real settings. In particular, we plan to incorporate automated labeling and semi supervised approaches for better scaling of the problem.
\section{Discussion}
\label{sec:discussion}
From the results, it is evident that there exists an important role played by semantic augmentation in the process of detection of assembly level vulnerability in graphs. Although previous studies involving graph neural networks were able to prove the ability of structural representations of programs to enhance vulnerability classification, the findings presented above reveal that structural information alone may not be adequate for proper analysis of binary codes. The vulnerability and non-vulnerability functions used in the experiment have been observed to behave similarly in terms of control flow structures especially after compiler optimization.

The SEMA-GUARD method employed here addresses the problem mentioned above by augmenting graph representations with lightweight semantic reasoning. Instead of looking at instructions as mere symbols, the framework tries to model some behavioral features such as stack operation, memory access, taint propagation and indirect control flow. Such semantically augmented graph representations allow a deeper understanding of how instructions interact when executed rather than just analyzing their structure. It is notable to mention that a significant number of vulnerabilities are a result of improper execution semantic of functions rather than any structural flaw.

As can be seen, semantic feature integration provides a better classification stability than purely statistical and/or purely structural baselines. First of all, traditional models based on opcodes and frequencies do not consider contextual information since they treat instructions separately and ignore their execution order. Secondly, CFG-based approaches take into account only structure without any distinction between different execution behaviors that could happen in the same graph. Our approach addresses this issue by including contextually sensitive information into graph representation prior to graph learning.

One of the conclusions that we make based on the experiment is that behavior-oriented abstraction might prove more useful for detecting vulnerabilities in binaries than purely syntactic one. For example, different optimizations like register allocation, instruction scheduling, inlining, loop optimizations, and code generation specific to certain architectures can significantly change the sequence of instructions in assembly code without changing its functionality. Consequently, approaches that rely on strict syntax and specific constructions introduced by particular compilers will generalize poorly when it comes to analyzing binaries from other software ecosystems.

The findings additionally suggest that graph neural networks remain well suited for binary vulnerability analysis because they preserve relational context between instructions. Unlike sequence-based models that process instructions linearly, graph-based representations capture branching behavior, loop structures, and execution dependencies more naturally. This capability is particularly important in low level security analysis, where vulnerabilities frequently emerge from interactions between multiple execution paths rather than isolated instructions. By combining semantic feature engineering with graph message passing, the proposed framework attempts to model both local instruction behavior and broader execution context simultaneously.

Another important observation relates to the connection between semantic reasoning and explainability. One of the most common critiques that apply to deep learning solutions in cybersecurity is their lack of explainability. Neural pipelines used for binary classifications typically do not explain much about how and why an artifact should be classified as vulnerable. On the other hand, semantic representations enable at least partially explainable abstractions since critical sections could be linked to specific behaviors such as unsafe stack writing, indirect jumps, suspicious memory accesses, or taint flow. The current pipeline does not provide comprehensive semantic explanation but paves a way for developing more explainable frameworks.

Speaking about practical applications of the approach in cyber defense, binary-based vulnerability detection still plays an important role. Often, cybersecurity professionals have to deal with programs for which source code is missing or unavailable. This situation occurs when one analyzes firmware, industrial control systems, embedded platforms, malware, etc. Under such circumstances, it is important to detect vulnerabilities in binaries despite a lack of information. Thus, techniques able to identify and analyze semantic concepts could prove to be helpful in practical cybersecurity applications.

On the other hand, there are several important issues left unsolved by this study in the domain of graph based vulnerability analysis. First of all, there is the issue of compilers' differences. In particular, different compiler families and optimization levels can produce assembly representations that significantly differ from one another despite having the same source-level semantics. This poses problems due to possible distribution shifts that may be detrimental to machine learning generalization. While the suggested approach utilizes semantic abstractions that should decrease compiler sensitivity, the experiments carried out in the paper do not properly solve this issue in full generality.

Another unresolved problem relates to interprocedural reasoning. Currently, the framework analyzes vulnerabilities in terms of semantic graph behavior only at the level of single functions. In practice, however, most vulnerable software is vulnerable because vulnerabilities appear due to interdependencies between different procedures or even different modules. For example, pointer vulnerabilities arise due to the propagation of pointers along the call chain; authentication bypass is usually performed using different parts of the program; and many memory vulnerabilities are caused by indirect execution flows.

It is worth noting that another aspect discussed in the present paper concerns the need for realistic data in vulnerability detection studies. On one hand, it is extremely convenient to have benchmark datasets like Juliet Test Suite because they allow researchers to conduct well controlled experiments. However, benchmarking has limitations since real datasets are much more diverse and heterogeneous than simulated ones. Indeed, real software products contain all kinds of optimizations; the code may be processed by different compilers, and the vulnerabilities may be very unbalanced in distribution among other features. Thus, shifting from benchmarking to actual vulnerable and fixed datasets is a crucial step for the research.

Finally, yet another implication of the work under consideration refers to the intersection of symbolic reasoning and machine learning as applied to the field of cybersecurity. Static analysis, whether through symbolic execution or taint analysis, implies strong semantic precision; however, it is usually limited by its scalability. By contrast, deep learning solutions are scalable and capable of representation learning, yet their semantic interpretability remains questionable. The present approach can be regarded as hybrid in the sense that semantics is combined with scalable graph learning.

In summary, the results obtained indicate that the direction of vulnerability detection based on semantic graphs is a promising one to pursue. It becomes apparent that by integrating execution-aware semantics in the design of graph-based neural networks, improved classification of software vulnerabilities can be achieved relative to statistical and structural methods alone. However, most importantly, it seems that future advancements in binary vulnerability detection cannot be solely reliant on neural network architecture; instead, the use of semantics must also be taken into account.
\section{Limitations}
\label{sec:limitations}
Nevertheless, there are a number of drawbacks that must be pointed out regarding the aforementioned technique.

For one thing, the use of Juliet Test Suite data set is questionable, since the programs it comprises of have been generated synthetically by means of implementing specific vulnerability patterns. This dataset might not fully represent the diversity of programs available in the real world.

Secondly, only non Windows test cases could be considered because of the lack of necessary tools and resources to conduct analysis in this particular environment. As such, the evaluation might have been biased due to the limited scope of application.

Another limitation is the absence of compiler diversity evaluation. Since different compilers and optimization levels may alter assembly structure significantly, future work should investigate cross compiler robustness.

In addition, the present framework is built around a function analysis approach. In reality, a variety of vulnerabilities may emerge through the interaction of multiple functions.

Lastly, while modeling the stack and propagation through it in order to analyze the semantics of program execution proves to be quite effective, there still remain various aspects of a program's behavior that cannot be accurately modeled using simplified heuristics.

\section{Conclusion}
\label{sec:conclusion}
SEMA-GUARD was introduced as a methodology in the paper, which uses semantic analysis and Graph Neural Networks to detect potential vulnerabilities in assembly code. This methodology can be employed in scenarios when source code is not available and, thus, it should extract the necessary information from the compiled version. Using different signals related to both the control flow and low level behaviors, such as using the stack and memory, and tainting, the methodology captures different sides of vulnerability detection tasks.

Experimental evaluation using a dataset generated from the Juliet Test Suite \cite{juliet} yielded the accuracy and F1-scores of 85.1\% and 0.801, respectively. As opposed to the baselines used in the experiments, the proposed solution consistently outperforms the baselines across all evaluated metrics. Specifically, the proposed method decreases false positives compared to pattern based solutions while maintaining high recall rates compared to graph-based models without semantic features.

However, further analysis suggests that the solution effectively detects typical vulnerabilities related to memory safety violations and control-flow problems. Nevertheless, it fails to catch some of the existing vulnerabilities due to the lack of consideration of the interactions between various instructions. Overall, the findings demonstrate that integrating semantic reasoning with graph-based representations provides a practical and effective direction for assembly-level vulnerability detection. By operating directly on compiled code, SEMA-GUARD addresses an important gap in binary analysis scenarios where source code is unavailable. The proposed framework establishes a foundation for future research on scalable and semantically informed binary security analysis.

From an application perspective, the approach can be applied for use in binary analysis problems such as firmware analysis, malware triaging, and audits of third party libraries. Since the technique works directly on assembly code, it can easily be incorporated into current reverse engineering processes without the need to access source code. Also, since SEMA-GUARD has a modular architecture, the individual modules such as semantic feature extraction or graph modeling can be customized to fit different application domains.

Some of the future work opportunities that arise from this experiment include considering inter-procedural dependencies by analyzing beyond function boundaries in order to detect more advanced vulnerabilities. Additionally, the introduction of more complex semantic models such as symbolic reasoning and enhanced data flow analysis would be useful to minimize the number of false negatives. Furthermore, testing the effectiveness of SEMA-GUARD on real life binary datasets would be beneficial. Finally, experimenting with different GNN architectures and training methodologies could be explored.

In conclusion, this study demonstrates that combining semantics and graphs in vulnerability detection is a viable approach, consistent with some of the recent advancements in graph based vulnerability detection approaches \cite{coca2024}.

\section*{Declarations}

\begin{itemize}
\item Conflict of Interest: Authors do not have any conflict of interest
\item Data and Code Avaliability: To support reproducibility and future research, the dataset and the code, preprocessing scripts, and baseline implementations are publicly available through Zenodo repositories. The dataset and implementation are publicly available at: \url{https://doi.org/10.5281/zenodo.20146077}

\item Funding: No funding has been used

\item Ethics approval and consent to participate: Not applicable
\item Consent for publication: Authors have consent for publication with subscription.
\item Materials availability: Not applicable
\item Code availability: \url{https://github.com/dursunoglu/SEMAGUARD}
\item Author contribution: Halil Ibrahim Dursunoglu was responsible final draft, experiments and result analysis. Kaan Sulkalar and Halil Ibrahim Dursunoglu implemented the code together. Kaan Sulkalar wrote the first draft.

\end{itemize}


\end{document}